# STRONG RF FOCUSING FOR LUMINOSITY INCREASE: SHORT BUNCHES AT THE IP

A. Gallo, P. Raimondi, M. Zobov, INFN - LNF

*Abstract*

One of the key-issues to increase the luminosity in the next generation particle factories is to reduce the bunch length at the interaction point (IP) as much as possible. This will allow reducing proportionally the transverse beta functions at the IP and increasing the luminosity by the same factor. The strong RF focusing consists in obtaining short bunches by substantially increasing the lattice momentum compaction and the RF gradient. In this regime the bunch length is modulated along the ring and could be minimized at the IP. If the principal impedance generating elements of the ring are located where the bunch is long (in the RF cavities region) it is possible to avoid microwave instability and excessive bunch lengthening due to the potential well distortion.

## 1. INTRODUCTION

The minimum value of the vertical beta-function $\beta_y$ at the IP in a collider is set by the hourglass effect [1] and it is almost equal to the bunch length $\sigma_z$. Reduction of the bunch length is an obvious approach to increase the luminosity. By scaling the horizontal and vertical beta functions $\beta_x$ and $\beta_y$ at the IP as the bunch length $\sigma_z$, the linear tune shift parameters $\xi_{x,y}$ remain unchanged while the luminosity scales as $1/\sigma_z$ [2]:

$$L \propto \frac{1}{\sigma_x \sigma_y} \propto \frac{1}{\sqrt{\beta_x \beta_y}} \propto \frac{1}{\sigma_z} \qquad (1)$$

A natural way to decrease the bunch length is to decrease the storage ring momentum compaction and/or to increase the RF voltage. However, in such a way we cannot obtain very short bunches since the short-range wakefields prevent this because of the potential well distortion and microwave instability.

In this paper we consider an alternative strategy to get short bunches at the IP. In particular, we propose to use strong RF focusing [3] (i.e. high RF voltage and high momentum compaction) to obtain very short bunches at the IP with progressive bunch elongation toward the RF cavity.

With respect to the case of short bunches with constant length all along the ring, the situation seems more comfortable since the average charge density driving the Touschek scattering is smaller. Besides, this allows placing the most important impedance generating devices near the RF cavity where the bunch is longest thus minimizing the effect of the wakefields.

## 2. STRONG RF FOCUSING

In order to compress the bunch at the IP in a collider a strong RF focusing can be applied. For this purpose high values of the momentum compaction factor $\alpha_c$ and extremely high values of the RF gradient are required. It is estimated that, for a $\Phi$-factory collider, an RF voltage $V_{RF}$ of the order of 10 MV is necessary provided that the $\alpha_c$ value is of the order of 0.2.

Under these conditions the synchrotron tune $\nu_s$ grows to values larger than 0.1 and the commonly used "smooth approximation" in the analysis of the longitudinal dynamics is no longer valid. Instead, the longitudinal dynamics is much more similar to the transverse one, and can be analyzed on the base of transfer matrices of the simple linear model reported in Fig. 1. In this model the cavity behaves like a thin focusing lens in the longitudinal phase space, while the rest of the machine is a drift space, where the "drifting" variable is the $R_{56}(s)$. In Fig. 1 $\lambda_{RF} = c/f_{RF}$ is the RF wavelength, $E/e$ is the particle energy in voltage units, while $L$ is the total ring length.

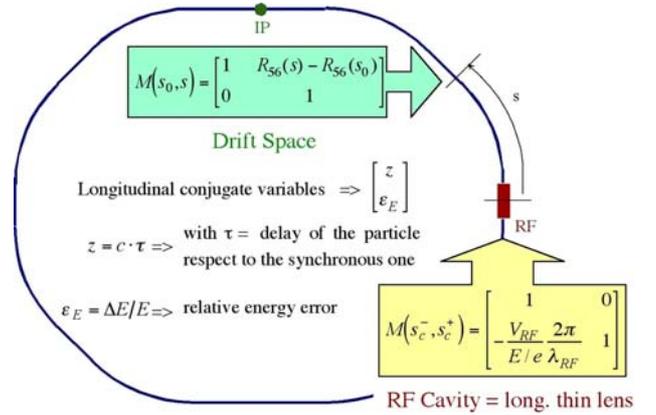

Figure 1: Longitudinal dynamics linear model.

The $R_{56}(s)$ parameter relates the path length to the normalized energy deviation of a particle, and is given by:

$$R_{56}(s) = \int_0^s \frac{\eta(\tilde{s})}{\rho(\tilde{s})} \, d\tilde{s} \qquad (2)$$

where $\rho(s)$ is the local bending radius and $\eta(s)$ is the ring dispersion function.

Taking the cavity position as the reference point $s = 0$, the one-turn transfer matrix $M(s, s+L)$ of this system

starting from the generic azimuth $s$ is given by:

$$M(s,s+L) = \begin{bmatrix} 1 - 2\pi \dfrac{R_{56}(s)}{\lambda_{RF}} \dfrac{V_{RF}}{E/e} & \alpha_c L \left(1 - 2\pi \dfrac{R_{56}(s)}{\lambda_{RF}} \left(1 - \dfrac{R_{56}(s)}{\alpha_c L}\right) \dfrac{V_{RF}}{E/e}\right) \\ -\dfrac{V_{RF}}{E/e} \dfrac{2\pi}{\lambda_{RF}} & 1 + 2\pi \dfrac{R_{56}(s)}{\lambda_{RF}} \left(1 - \dfrac{\alpha_c L}{R_{56}(s)}\right) \dfrac{V_{RF}}{E/e} \end{bmatrix} \quad (3)$$

The one turn synchrotron phase advance is given by:

$$\cos\mu = \frac{1}{2}Tr[M(s,s+L)] = 1 - \pi \frac{\alpha_c L}{\lambda_{RF}} \frac{V_{RF}}{E/e} \quad (4)$$

leading to the following stability condition:

$$|\cos\mu| \leq 1 \Rightarrow \mu \leq \pi \Rightarrow \nu_s \leq 1/2 \Rightarrow$$
$$\Rightarrow V_{RF} \leq \frac{2}{\pi} \frac{\lambda_{RF}}{\alpha_c L} E/e = V_{RF_{Max}} \quad (5)$$

which shows that there is a constraint in the choice of the values of $V_{RF}$ and $\alpha_c$. The one-turn transfer matrix can be put in the canonical form:

$$M(s,s+L) = \cos\mu \cdot \hat{I} + \sin\mu \cdot \hat{J} =$$
$$= \cos\mu \cdot \begin{bmatrix} 1 & 0 \\ 0 & 1 \end{bmatrix} + \sin\mu \cdot \begin{bmatrix} \alpha_l & \beta_l \\ -\gamma_l & -\alpha_l \end{bmatrix} \quad (6)$$

and the longitudinal Twiss parameters are given by:

$$\alpha_l(s) = \frac{1-\cos\mu}{\sin\mu}\left[1 - \frac{2R_{56}(s)}{\alpha_c L}\right]$$
$$\beta_l(s) = \frac{\alpha_c L}{\sin\mu}\left[1 - (1-\cos\mu)\frac{2R_{56}(s)}{\alpha_c L}\left(1 - \frac{R_{56}(s)}{\alpha_c L}\right)\right] \quad (7)$$
$$\gamma_l(s) = \frac{1-\cos\mu}{\sin\mu}\frac{2}{\alpha_c L}$$

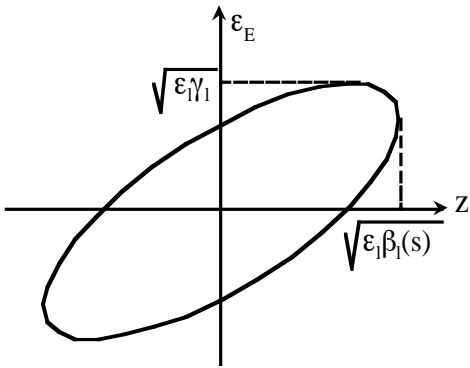

Since $\gamma_l$ does not depend upon $s$, the vertical size of the ellipse (which represent the normalized energy spread $\sigma_E/E$ of the equilibrium distribution) does not vary along the ring.

The longitudinal emittance $\varepsilon_l$ is related to the equilibrium energy spread according to:

$$\sigma_E/E = \sqrt{\varepsilon_l \gamma_l} \Rightarrow \varepsilon_l = (\sigma_E/E)^2 \frac{\sin\mu}{2\pi} \frac{E/e}{V_{RF}}\lambda_{RF} =$$
$$= (\sigma_E/E)^2 \frac{\sin\mu}{1-\cos\mu} \frac{\alpha_c L}{2} \quad (8)$$

On the contrary, since $\beta_l$ does depend upon $s$, the horizontal size of the ellipse (i.e. the bunch length $\sigma_z$) varies along the ring according to:

$$\sigma_z(s) = \sqrt{\varepsilon_l \beta_l(s)} = (\sigma_E/E) \cdot$$
$$\cdot \sqrt{\frac{\alpha_c L}{2\pi} \frac{E/e}{V_{RF}}\lambda_{RF}\left(1 - 2\pi \frac{R_{56}(s)}{\lambda_{RF}}\left(1 - \frac{R_{56}(s)}{\alpha_c L}\right)\frac{V_{RF}}{E/e}\right)} = \quad (9)$$
$$= \sigma_z(0) \sqrt{1 - 2\pi \frac{R_{56}(s)}{\lambda_{RF}}\left(1 - \frac{R_{56}(s)}{\alpha_c L}\right)\frac{V_{RF}}{E/e}}$$

where $\sigma_z(0)$ is the bunch length at $s=0$ (i.e. at the cavity position). It may be noticed that $\sigma_z(0) = \sigma_{z_{max}}$ is the maximum value of the bunch length along the ring. On the other hand, the minimum value $\sigma_{z_{min}}$ corresponds to the $s_{min}$ position where $R_{56}(s_{min}) = \alpha_c L/2$. If the position of the minimum corresponds to the IP one gets:

$$\sigma_z(IP) = \sigma_z(Cav) \sqrt{1 - \frac{\pi}{2}\frac{\alpha_c L}{\lambda_{RF}}\frac{V_{RF}}{E/e}} =$$
$$= \sigma_z(Cav) \sqrt{\frac{1+\cos\mu}{2}} \quad (10)$$

As $\mu$ approaches $180°$, the ratio between the bunch lengths at the IP and at the RF goes to zero. This result is of great interest since it allows designing a ring where the bunch is short at the IP and progressively elongates moving toward the RF position.

## 3. EQUILIBRIUM ENERGY SPREAD

In order to compute exactly the bunch size along the ring one needs to know the longitudinal emittance value (or, equivalently, the value of the equilibrium energy spread). These values can be worked out from a rigorous analysis of the longitudinal dynamics (abandoning the smooth approximation) or from a multi-particle tracking simulation including the distributed emission process along the machine. We follow an analytical approach based on the computation of the second momenta of the bunch equilibrium distributions using the eigenvectors of the longitudinal one-turn transfer matrix [4] that gives the following result:

$$\left(\frac{\sigma_E}{E}\right)^2 = \frac{1}{1+\cos\mu} \frac{55}{48\sqrt{3}} \frac{r_e \hbar \gamma^5 \tau_d}{m_e L} \cdot \oint \left[1 - \frac{2\pi R_{56}(s)}{\lambda_{RF}}\left(1 - \frac{R_{56}(s)}{\alpha_c L}\right)\frac{V_{RF}}{E/e}\right]\frac{ds}{|\rho(s)|^3} \quad (11)$$

where $r_e$ and $m_e$ are the electron classical radius and rest mass, $\tau_d$ is the longitudinal damping time and $\gamma = E/(m_e c^2)$ is the relativistic factor.

It may be noticed that the equilibrium energy spread $\frac{\sigma_E}{E}$ is diverging as $\mu$ tends to $\pi$, while at low tunes it tends to the value $\left.\frac{\sigma_E}{E}\right|_0$:

$$\left(\left.\frac{\sigma_E}{E}\right|_0\right)^2 = \frac{55}{96\sqrt{3}} \frac{r_e \hbar \gamma^5 \tau_d}{m_e L} \oint \frac{ds}{|\rho(s)|^3} \quad (12)$$

which is the expression reported in literature [5].

Expression (12) can be also conveniently rewritten in the following forms:

$$\left(\frac{\sigma_E}{E}\right)^2 = \frac{1}{1+\cos\mu} \frac{55}{48\sqrt{3}} \frac{r_e \hbar \gamma^5 \tau_d}{m_e L} \oint \frac{\beta_l(s)}{\beta_l(0)|\rho(s)|^3} ds = \\ = \left(\left.\frac{\sigma_E}{E}\right|_0\right)^2 \frac{2}{1+\cos\mu} \frac{\oint \frac{\beta_l(s)}{\beta_l(0)|\rho(s)|^3} ds}{\oint \frac{ds}{|\rho(s)|^3}} \quad (13)$$

In the simplified assumption of constant bending radius $\rho$ and $R_{56}(s)$ linearly growing in the arcs, expression (12) reduces to:

$$\left(\frac{\sigma_E}{E}\right)^2 = \frac{2}{3}\left(\left.\frac{\sigma_E}{E}\right|_0\right)^2 \frac{2+\cos\mu}{1+\cos\mu} \quad (14)$$

Different results may be obtained if the ring has variable bending radii and/or the $R_{56}(s)$ function does not grow linearly in the arcs.

Under the assumptions leading to eq. (14), the longitudinal emittance $\varepsilon_l$ and the bunch lengths at the RF cavity and IP are given by:

$$\varepsilon_l = \frac{\alpha_c L}{3}\left(\left.\frac{\sigma_E}{E}\right|_0\right)^2 \frac{2+\cos\mu}{\sin\mu};$$

$$\sigma_z(Cav) = \frac{\alpha_c L}{\sin\mu}\left(\left.\frac{\sigma_E}{E}\right|_0\right)\sqrt{\frac{2+\cos\mu}{3}}; \quad (15)$$

$$\sigma_z(IP) = \alpha_c L\left(\left.\frac{\sigma_E}{E}\right|_0\right)\sqrt{\frac{2+\cos\mu}{6(1-\cos\mu)}}$$

The emittance and the bunch length at the RF cavity, as well as the energy spread, diverge as $\mu$ approaches $\pi$, while the bunch length at the IP remains finite

Figure 2 shows the longitudinal emittance and the equilibrium energy spread as a function of the phase advance $\mu$. The lines correspond to the analytical expressions (14)-(15), while dots represent the results of the multi-particle tracking simulations. The bunch length dependences on $\mu$ (both analytical and numerical) are reported in Fig. 3.

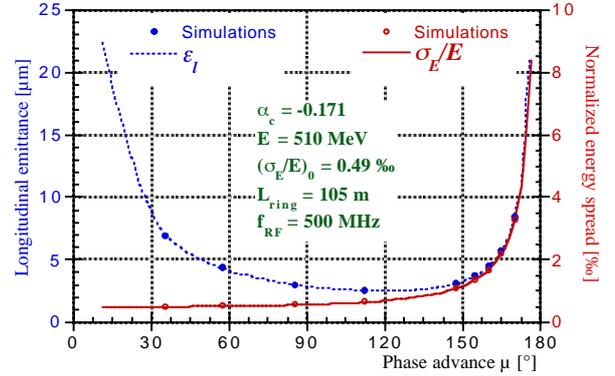

Figure 2: Longitudinal emittance and energy spread vs. phase advance

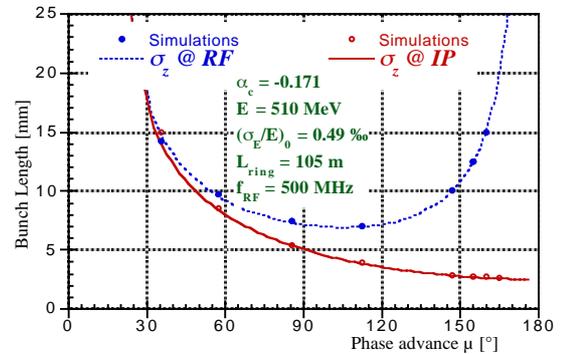

Figure 3: Bunch length @ RF and IP vs. phase advance

As it is seen, the longitudinal emittance exhibits a minimum at $\mu = 120°$. The analogy with the transverse case is quite evident [6]. Being the momentum compaction fixed, the various phase advances correspond to different values of the RF voltages. The voltage required to approach the limit phase advance value of 180° exceeds 10 MV. The use of superconducting cavities is mandatory in this case.

## 4. RF ACCEPTANCE

In a storage ring the RF acceptance is defined as the maximum energy deviation corresponding to a stable particle trajectory in the longitudinal phase space. The trajectory associated to the maximum energy deviation is the so-called "separatrix" since it separates the stable (closed) to the unstable (open) trajectories in the longitudinal phase space.

In a standard, low $\nu_s$ storage ring the longitudinal phase space trajectories are independent on the particular observation abscissa $s$. The RF acceptance is computed as the separatrix half-height [5].

In the strong RF focusing case, the longitudinal phase space trajectories configuration changes along the ring, and the same does the separatrix, as shown in Figure 4.

In order to compute the particle loss rate caused by the Touschek scattering process, the relevant RF acceptance is the half-heigth of the separatrix section at $z=0$, which is a function of the azimuth $s$ in this case. In fact, the maximum acceptable energy variation for a particle starting from about the origin of the phase space is smaller in the RF cavity region (where the bunch is longer); this must be taken into account in lifetime evaluations.

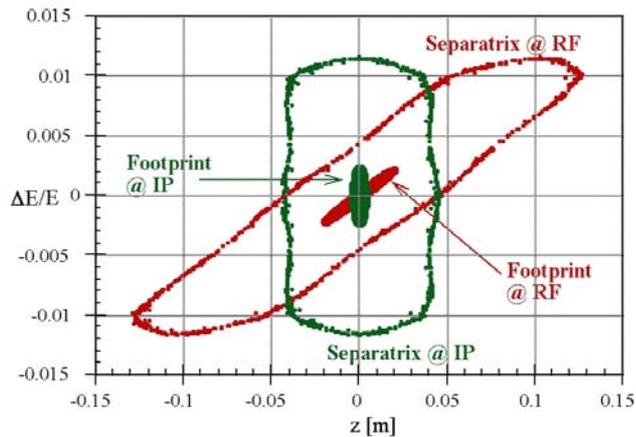

Figure 4: Separatrices in a strong focusing storage ring

Concerning the loss rate caused by statistical fluctuations of the particle energy deviation (quantum lifetime) the relevant parameter is the separatrix area, which is independent on the azimuth $s$ as a consequence of the Liouville theorem.

## 5. BUNCH LENGTHENING

In a collider based on the strong RF focusing concept the bunch at the IP has to remain short up to the design bunch current value. This is a critical point, since the effect of the machine wakefields is strongly dependent on the bunch length. In the case of pure inductive impedance, for example, the wake potential scales as $1/\sigma_z^2$. On the other hand, the strong RF focusing produces a modulation of the bunch length along the machine. To minimize the bunch lengthening effect it is worth to design a storage ring with all the most dangerous impedance generating components (such as injection kickers, longitudinal and transverse kickers of the fast feedback systems, monitors and striplines, …) placed where the bunch is longer, i.e. near the RF cavity section.

To estimate the criticality of the impedance location we have performed multi-particle tracking simulation of a bunch in the strong RF focusing regime using the DAΦNE wake (see, for example, [7]) in the two extreme (and unrealistic) cases of impedance completely concentred near the RF or near the IP.

In the first case the simulation results, reported in Fig. 5, show that with the chosen parameters ($\alpha_c = -.17$, $\mu = 165°$, $3 \cdot 10^{10}$ particles $\equiv I_b = 15$ mA) the bunch is about 3 mm long at the IP, and 11 mm long at the RF location, with no significant degradation with respect to the bunch length at zero current.

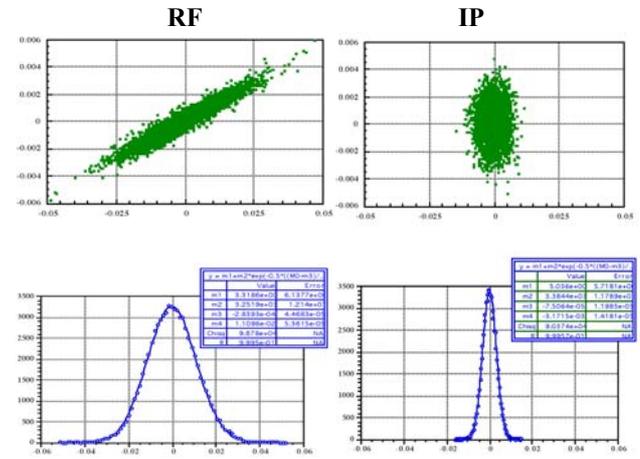

Figure 5: Particle distribution at RF and IP position in the case of a wake located near the RF section

In the second case, where the wake is concentred near the IP, the results of the tracking simulations are reported in Figure 6 and show a strongly deformed bunch profile as a consequence of a microwave instability.

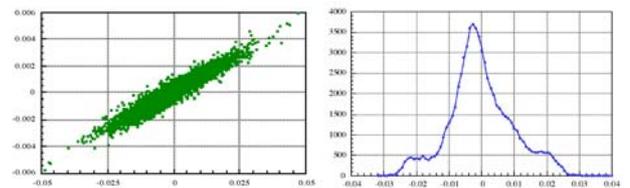

Figure 6: Particle distribution at RF position in the case of a wake located near the IP section

We have also considered the case of positive momentum compaction ($\alpha_c = +0.17$) and wake concentred near the RF. The simulations show that the bunch is stable with a very moderate lengthening.

## 6. CONCLUSIONS

In this paper the motion of particles in a strong longitudinal focusing storage ring is described by means of the linear matrices formalism. Longitudinal optical functions are derived, showing that the bunch length varies along the ring and may be minimized at the IP.

Analytical expressions for the longitudinal emittance and the energy spread of the bunch equilibrium distribution have been obtained and validated by comparison with results from multiparticle tracking simulations. It has been shown that the longitudinal emittance and the energy spread, as well as the bunch length at the RF cavity position diverge as the synchrotron phase advance approaches 180° per turn, while the bunch length at the IP tends to a minimum value which is finite.

Many aspects of beam physics need to be studied to establish whether or not a collider may efficiently work in the strong longitudinal focusing regime. The most relevant issues are bunch lengthening due to the wakefields, Touschek lifetime, dynamic aperture and beam-beam effect. Very preliminary multiparticle tracking simulations based on the DAΦNE short range wake show that the short bunch length at the IP can be preserved up to relatively high bunch current (> 10 mA) provided that all the wake is concentrated near the RF cavity, the position where the bunch is longest.